\documentclass[conference]{IEEEtran}
\IEEEoverridecommandlockouts
\usepackage{cite}
\usepackage{amsmath,amssymb,amsfonts}
\usepackage{algorithmic}
\usepackage{graphicx}
\usepackage{textcomp}
\usepackage{xcolor}
\def\BibTeX{{\rm B\kern-.05em{\sc i\kern-.025em b}\kern-.08em
    T\kern-.1667em\lower.7ex\hbox{E}\kern-.125emX}}

\usepackage{subfig}
\usepackage{scalerel}
\DeclareMathOperator*{\concat}{\scalerel*{\Vert}{\sum}}
\usepackage{multirow}
\usepackage{amssymb}
\usepackage{booktabs}
\usepackage{float}
\usepackage{hyperref}

\begin{document}

\title{Improving Domain Generalization for Sound Classification with Sparse Frequency-Regularized Transformer}

\author{
\IEEEauthorblockN{Honglin Mu, Wentian Xia, Wanxiang Che}
\IEEEauthorblockA{\textit{Research Center for Social Computing and Information Retrieval}}
Harbin Institute of Technology, China \\
\{hlmu, wtxia, car\}@ir.hit.edu.cn
}

\maketitle

\begin{abstract}
Sound classification models' performance suffers from generalizing on out-of-distribution (OOD) data. Numerous methods have been proposed to help the model generalize. However, most either introduce inference overheads or focus on long-lasting CNN-variants, while Transformers has been proven to outperform CNNs on numerous natural language processing and computer vision tasks. We propose FRITO, an effective regularization technique on Transformer's self-attention, to improve the model's generalization ability by limiting each sequence position's attention receptive field along the frequency dimension on the spectrogram. Experiments show that our method helps Transformer models achieve SOTA generalization performance on TAU 2020 and Nsynth datasets while saving 20\% inference time.

\end{abstract}


\section{Introduction}
\label{sec:intro}


Machine learning has made significant strides in the realm of sound classification. However, factors such as variable ambience and device-specific sound distortion~\cite{dcase2019sound} often impede models that excel on the training set from generalizing effectively upon deployment. This challenge has garnered substantial attention, as evidenced by the DCASE Challenges~\cite{DCASE2019Workshop, DCASE2020Workshop, DCASE2021Workshop, DCASE2022Workshop}. These competitions introduce a cross-device acoustic scene classification (ASC) task, where models are trained to discriminate the scene of the input fragment, e.g., metro or square, while recording devices differ between the training set and the test set. Performance on the unseen devices (i.e., domains) within the test set serves as an indicator of the model's generalization capabilities.

\begin{figure}[h]
\centering
\includegraphics[width=\linewidth]{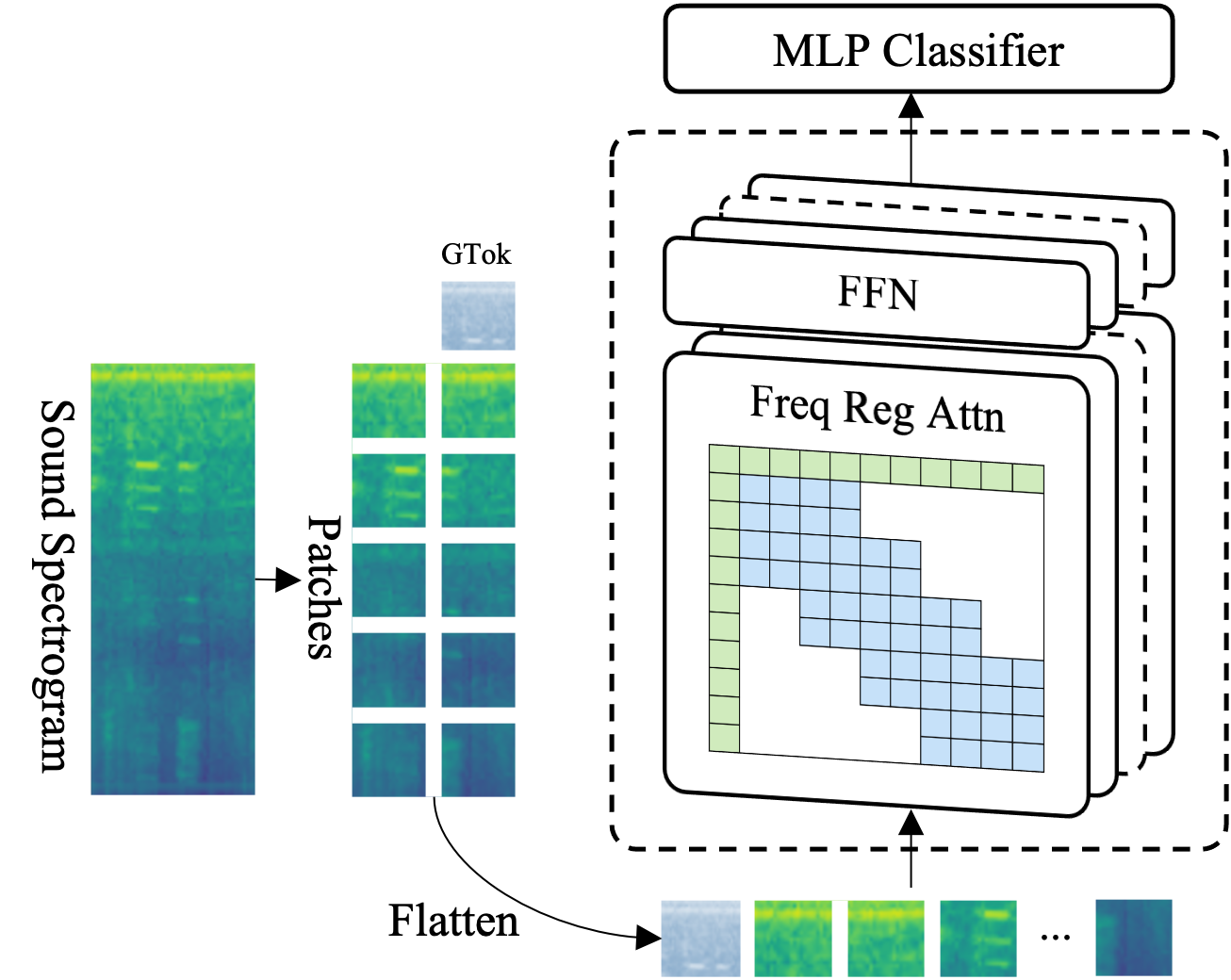}
\caption{The Frequency Regularized Transformer. \textit{GTok} is abbreviation for global token. }
\label{fig:arch}
\end{figure}

We investigated domain generalization methods in DCASE 2019-2022, finding that most sound classification generalization methods, despite improve models' generalization ability, either introduce overheads~\cite{mukhoti2020calibrating, suh2020designing, hu2021two, shim2022attentive} or based on durable model structures such of Resnets~\cite{eghbal2019acoustic, mcdonnell2020acoustic, kim2022qti, lee2022hyu}. We focus on the newer model structure optimized for domain generalization.

CNN-variants, such as VGG~\cite{simonyan2014very}, Resnet~\cite{he2016deep}, and FCNN~\cite{hu2021two}, are de facto domination backbones in previous work.
Koutini et al.~\cite{koutini2019receptive} found that CNNs suffer from overfitting on biased data in the ASC task, which could be alleviated by restricting CNNs' receptive field.
Similarly, McDonnell et al.~\cite{mcdonnell2020acoustic} bolstered CNN's performance on OOD devices by limiting the model's perception of frequency. The proposed CNN variant includes two separate inference paths, each probing half range on the spectrogram's frequency and fusing with the other before the classification head. 
With Transformer~\cite{vaswani2017attention} applied to the text~\cite{devlin2018bert, liu2019roberta, brown2020language}, image~\cite{dosovitskiy2020image} and speech~\cite{baevski2020wav2vec, chen2022wavlm}, its power in the sound classification has also been discovered~\cite{koutini2021efficient, gong2021ast}; the Patchout faSt Spectrogram Transformer (PaSST)~\cite{koutini2021efficient} surpassed CNN variants in performance and efficiency through dropping out patches from the spectrogram. While Transformer-based models challenge sound classification tasks, their generalization degradation issue remains to be investigated.

This work focuses on improving the Transformer's domain generalization ability in sound classification tasks. Inspired by Koutini et al. and McDonnell et al., we propose the Frequency-Regularized Transformer which fuses frequency regularization into self-attention to provide both robustness and efficiency. We limit the Transformer's receptive field along the frequency on the sound's spectrogram by applying two types of masks to the Transformer's self-attention, one of which perceives neighbor frequencies while the other considers the entire range. Such a strategy alleviate the over-fitting problem that the model suffers facing OOD data. Our contributions can be summarized as follows:

\begin{itemize}
    \item We propose the \textbf{F}requency-\textbf{R}egular\textbf{I}zed \textbf{T}ransf\textbf{O}rmer (FRITO), a Transformer variant enhanced on the generalization ability for sound classification, which achieves SOTA on unseen domains of TAU 2020~\cite{Mesaros2018_DCASE} and Nsynth~\cite{engel2017neural} datasets.
    \item We implement the sparse form of this method's common use case, saving 20\% of inference time and 21\% memory compared to its full-attention version.
\end{itemize}

Our code for this paper will be publicly available at \url{https://github.com/hlmu/FRITO}.

\section{Frequency-Regularized Transformer (FRITO)}
\subsection{Backbone}
Our proposed method utilizes the Vision Transformer (ViT)~\cite{dosovitskiy2020image} as its backbone. ViT is a state-of-the-art deep learning model that has adapted the Transformer architecture for computer vision tasks. It achieves this by dividing the image into a fixed grid of patches, which are then added with positional encoding and fed into a standard Transformer block. This enables ViT to capture long-range dependencies between image patches and achieve impressive results on various vision tasks.

Building on this foundation, the PaSST model~\cite{koutini2021efficient} leverages ViT in the audio domain by splitting the audio spectrogram into patches and employing disentangled time and frequency positional encoding. In our approach, we take inspiration from both ViT and PaSST and modify the attention block to restrict its receptive field on the frequency dimension. This approach enables our model to mitigate overfitting issues and generalize better to out-of-domain data.

Our model architecture is illustrated in Figure~\ref{fig:arch}, and it is composed of modified ViT blocks that process the audio spectrogram patches.

\subsection{Frequency Attention Regularization}
In this section, we introduce our model's regularization scheme, which restricts the receptive field of the Transformer's each input position by adding masks to its self-attention. Our scheme includes local and global attention, similar to the efficient transformers~\cite{child2019generating, zaheer2020big}. This method differs from previous work in that it restricts the receptive field along the frequency dimension, aiming to improve the model's generalization ability instead of dealing with long sequences.

Given a piece of audio's Mel-spectrogram, we split it into patches and add positional encoding following PaSST. Let such patch matrix be $P\in \mathbb{R}^{h\times w\times d}$ where $h$,$w$, and $d$ represent rows of patches, columns of patches, and Transformer's hidden size, respectively. We then flatten the patch matrix into $x_{P(1,1)},\dots,x_{P(h,w)} \in \mathbb{R}^{hw\times d}$, and prefix $k$ global tokens $x_{g_1},\dots,x_{g_k}$ which will be explained shortly. Then the Transformer's input sequence can be represented as:
\begin{align}
X=(x_{g_1}^1,\dots,x_{g_k}^k,x_{P(1,1)}^{k+1},\dots,x_{P(h,w)}^{t}) \in \mathbb{R}^{t\times d} \nonumber 
\end{align}
where $t$ is the Transformer's input sequence length.

\begin{figure}%
    \centering
    \subfloat[$r=1$, $v=2$]{\label{subfig:attn1}{\includegraphics[width=.5\linewidth]{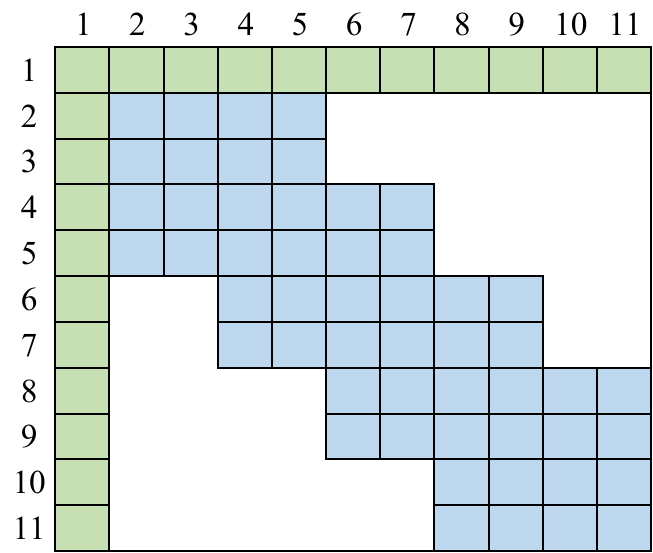} }}%
    \subfloat[$r=3$, $v=1$]{\label{subfig:attn2}{\includegraphics[width=.5\linewidth]{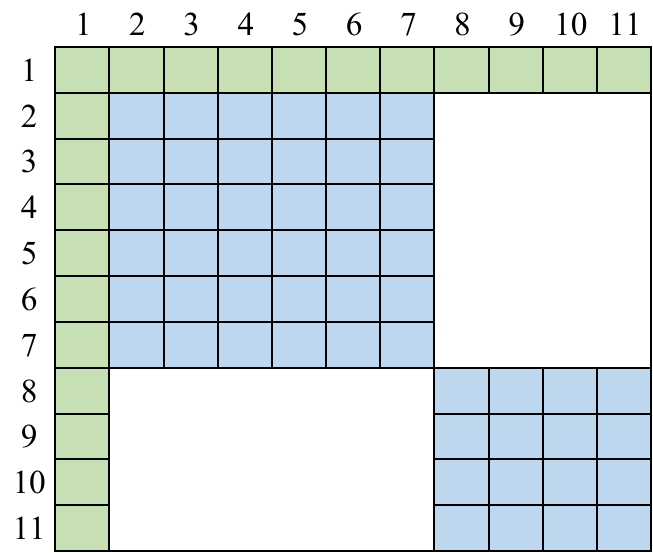} }}%
    \caption{Example attention masks $M$ for the input sequence, made from spectrogram patches with $5$ rows in frequency and $2$ columns in time, one global token prefixed. Blank and colored blocks represent $-\infty$ and $0$, respectively. Green blocks represent global attention, while the blue ones represent local attention. (a) Under the conditions of $r=1$ and $v=2$, directly adjacent rows are interiorly accessible. (b) When $r=3$ and $v=1$, the patches in the first three rows are visible to each other, and the patches in the last two rows are mutually sensible.}%
    \label{fig:attn_mat}%
\end{figure}

We constrain Transformer's receptive field by adding a mask matrix $M$ on its self-attention weights.
Let $Q, K, V\in~\mathbb{R}^{t\times d}$ be the Transformer's queries, keys, and values, respectively.
Then the attention of our Frequency-Regularized Transformer can be written as:
\begin{align}
ATTN_{FRITO}(X)=\text{softmax}(\frac{QK^T}{\sqrt{d}}+M)V
\label{eq:attn1}
\end{align}
where $M\in \{0,-\infty\}^{n\times n}$ is the attention mask matrix shown as~\ref{fig:attn_mat}. Sequence token $X_*^{j}$ is visible to $X_*^{i}$ iff $M(i,j)$ equals zero. The following sections provide a detailed design of the mask matrix $M$.

\subsection{Local Frequency Attention}
This subsection first presents how we select the receptive field of each token, and then introduces the $M$ matrix representation for this scheme. Specifically, we restrict the receptive field $S$ of a patch $P_{(a,b)}$ on the patch matrix $P$ to its neighbors close in frequency. The receptive field $S$ can be expressed using the following formula, where $a$ and $b$ represent arbitrary row and column numbers:

\begin{align*}
S(a,b)=\biggl\{(p, q) \Big|&p\in \Big[\bigl \lfloor\frac{a-1}{r}\bigr \rfloor r + 1, \bigl \lfloor\frac{a+r-1}{r}\bigr \rfloor rv \Big],\\
&q\in [1, w] \biggr\}
\end{align*}
where $r$ is the size of a row cluster with $r$ rows of internally visible patches; $v$ is the overlap factor, indicating the number of row clusters that $P_{(a,b)}$ can observe; the optimal value of $r$ and $v$ are chosen from experiments.

These receptive field rules are then applied to the attention mask $M$ for the flattened input sequence $X$, represented as:
\begin{align*}
M[&iw+k+1:(i+rv)w+k+1, \\
 &iw+k+1:(i+rv)w+k+1]=0, \\
 &i \in \{0,r,2r,\dots,\lfloor \frac{h-1}{r}\rfloor r \}
\end{align*}
where $k$ is the number of global tokens, which will be explained shortly. A visualized example is shown in Figure~\ref{fig:attn_mat}.

\subsection{Global Frequency Attention}
Despite local frequency insights, the model needs overall perception to perform the classification task. Similar to BIGBIRD-ETC's approach~\cite{zaheer2020big}, we add global tokens $x_{g_1},x_{g_2},\dots,x_{g_k}$ at the beginning of the flattened patch tokens to perceive the entire sequence. Global tokens share a mutual sight with all other tokens, and its attention mask $M$ can be expressed as:
\begin{align*}
M[i,:]=0, M[:,i]=0, i \in \{1, 2, \dots, k\}
\end{align*}
Values on other positions in $M$ not defined by local and global frequency attention are all $-\infty$. Figure~\ref{fig:attn_mat} visualize the $M$ matrix with an example.

\subsection{Sparse Frequency-Regularized Attention}
The above approach restricts the model's receptive field by adding a mask to the attention weight matrix. However, its na\"ive implementation does not improve the speed of reasoning. While the $M$ matrix has a relatively large number of $-\infty$ value, its sparsity can be utilized. This paper adopts a sparse attention operation for the case where $v=1$ to improve the model training and reasoning speed.

When $v=1$, the input sequence $X$ excluding the global tokens can be segmented into $l=\bigl\lceil \frac{h}{r} \bigl\rceil$ mutually imperceptible blocks denoted as $X_1, X_2, \dots, X_l$. The $Q, K, V$ in Equation~\ref{eq:attn1} can be split along the time dimension into $Q_i, V_i, K_i\in \mathbb{R}^{r\times d}, i\in{1,2,\dots,l}$, respectively, corresponding to the local attention of the $i$-th block. In this case, the attention formula in Equation~\ref{eq:attn1} can be rewritten as:
\begin{align*}
ATTN_{SLocal}(X)=\concat_{i=1}^l \text{softmax}\left( \frac{Q_iK_i^T}{\sqrt{d}}\right) V_i
\end{align*}
where $\mathbin\Vert$ represents concatenation along the time dimension.

The above formula only considers local attention, but FRITO also requires global attention. We aggregate the global tokens into a block represented as $X_g$, its query and key matrix denoted as $Q_g$ and $K_g$, respectively. The attention formula can then be rewritten as the final sparse version:
\begin{align*}
ATTN_{SFRITO}(X)=&\text{softmax}\left( \frac{Q_gK_g^T}{\sqrt{d}}\right) V \oplus \\
& \concat_{i=1}^l \text{softmax}\left( \frac{Q_i(K_g\mathbin\Vert K_i)^T}{\sqrt{d}}\right) V_i
\end{align*}
where both $\oplus$ and $\mathbin\Vert$ are concatenation operation.

Sparse FRITO is computationally equivalent to its full-attention version. In particular, since $Q$, $K$, and $V$ are only split along the time dimension, the weights obtained from the full-attention model can be directly used for sparse attention inference.

\section{Experiment Setup}

We evaluate our method on two publicly available datasets, TAU 2020~\cite{Mesaros2018_DCASE} and Nsynth~\cite{engel2017neural}.

\subsection{Nsynth}
The Nsynth dataset is a collection of three different sound sources, \textit{acoustic}, \textit{electronic}, and \textit{synthetic}, consisting of sounds from various instruments played at 16kHz, including \textit{bass}, \textit{brass}, \textit{flute}, \textit{guitar}, \textit{keyboard}, \textit{mallet}, \textit{organ}, \textit{reed}, \textit{string}, \textit{synth\_lead}, and \textit{vocal}. In addition, each instrument except \textit{synth\_lead} includes segments from \textit{acoustic} sound source.
We require models to infer on unseen \textit{electronic} and \textit{synthetic} source after training them to classify instruments on \textit{acoustic} source.
Original data split mixes \textit{synthetic}, \textit{acoustic}, and \textit{electronic} sources in train, development, and test set. While we aim to measure models' generalization ability on unseen domains, i.e. sound sources, we re-divide the train, development, and test set in the following way:

We retain all sounds from the \textit{acoustic} source, incorporating them into the training set, while utilizing \textit{synthetic} and \textit{electronic} sounds, with markedly different sound characteristics, as out-of-domain samples for the development set. The train-development proportion adheres to the original train-test ratio established in the original split. Detailed dataset split information can be found in Table~\ref{tab:nsynthsplit}.

\begin{table}[h]
\begin{center}
\caption{Statistics of the re-split Nsynth dataset}
\label{tab:nsynthsplit}
\begin{tabular}{l|c|ccc}
  \toprule
    \multirow{2}{*}{Instrument} & Train Set & \multicolumn{3}{c}{Dev Set} \\
  \cline{2-5}
     & Acoustic & Synthetic & Acoustic & Electronic \\
  \hline
    Bass & 180 & 792 & 20 & 110 \\
    Brass & 13740 & 0 & 20 & 70 \\
    Flute & 6552 & 104 & 20 & 35 \\
    Guitar & 13323 & 108 & 20 & 345 \\
    Keyboard & 8488 & 59 & 20 & 657 \\
    Mallet & 27702 & 107 & 20 & 341 \\
    Organ & 156 & 0 & 20 & 469 \\
    Reed & 14242 & 156 & 20 & 76 \\
    String & 20490 & 0 & 20 & 84 \\
    Vocal & 3905 & 121 & 20 & 140 \\
  \bottomrule
\end{tabular}
\end{center}
\end{table}

\subsection{TAU 2020}
The TAU 2020 dataset is directly used by DCASE~2020~Task~1A~\cite{DCASE2020Workshop} and DCASE~2021~Task~1A~\cite{DCASE2021Workshop} to evaluate the generalization properties of systems across a number of different devices. It is an environmental sound classification dataset that requires the model to classify the sound environment, e.g., \textit{metro}, \textit{shopping mall}, and \textit{public square}, of the input audio. The dataset is recorded by multiple recording devices. The development set contains devices with unique acoustic characteristics not presented in the training set, so the performance on it can reflect the model's generalization ability. We use this dataset following DCASE 2020's official instructions~\cite{dcase2020rule}.

\subsection{Model Initialization}
We use the pre-trained weight passt\_s\_swa\_p16\_128\_ap476 from the PaSST model, which has $8$ attention heads and a $768$-dimensional encoder, derived from DeiT-B$\uparrow 384$~\cite{touvron2021training}, finetuned on Audioset~\cite{gemmeke2017audio}.
For the PaSST experiment, we set \textit{s\_patchout\_t=10}, \textit{s\_patchout\_f=5} as in their work.
We finetune models on $4$ Tesla-A100-80GBs, on a batch size containing 800 seconds of sound, using the Adam optimizer with a maximum learning rate of $1e-3$.


\section{Results}

\subsection{Nsynth}

Our FRITO method demonstrates optimal overall performance on the Nsynth dataset, as shown in Table~\ref{tab:nsynth}. The dataset comprises three domains: \textit{acoustic}, \textit{electronic}, and \textit{synthetic}. We train models on the \textit{acoustic} domain, and evaluate their generalization performance on the remaining two domains. Consistent with previous work's experimental conclusions~\cite{koutini2021efficient}, Transformer-based methods outperform the Resnet overall. Among these Transformer-based methods, PaSST exhibits a degradation in the \textit{Synthetic} domain compared to ViT, while our method remains less affected.

\begin{table}[h]
\begin{center}
\caption{Accuracy on the Nsynth dataset. \textit{Electronic} and \textit{Synthetic} do not appear in the training set.}
\label{tab:nsynth}
\begin{tabular}{lccc}
  \toprule
  Method & Overall & Electronic & Synthetic \\
  \midrule
    Resnet~\cite{hu2021two} & 0.276 & 0.251 & \textbf{0.317} \\
    ViT~\cite{koutini2021efficient} & 0.301 & 0.351 & 0.221 \\ 
    PaSST~\cite{koutini2021efficient} & 0.285 & 0.355 & 0.173 \\ 
    FRITO-$r6$-$v1$ & \textbf{0.307} & \textbf{0.379} & 0.192 \\ 
  \bottomrule
\end{tabular}
\end{center}
\end{table}

\subsection{TAU 2020}

Performance on the TAU 2020 dataset is presented as Table~\ref{tab:tau2020}. Our regularized method achieved the optimal performance on the unseen domains $S4-S6$, with $3.5\%$ improvement compared to Resnet and $0.5\%$ improvement in generalization performance compared to PaSST. Transformer-based models in our experiments have significantly better generalization performance than Resnet, which is in accord with results on the Nsynth dataset.

\begin{table}[h]
\begin{center}
\caption{Accuracy on TAU 2020 development set. \textit{A, B, C} are real recording devices, while \textit{S1-S6} are virtual devices generated from \textit{A, B, C}. \textit{S4-S6} only appears in the development set. The results marked with a $\star$ symbol correspond to our own run.}
\label{tab:tau2020}
\begin{tabular}[H]{lcccc}
  \toprule
  Method & Overall & S4-S6 & S1-S3 & A \& B \& C \\
  \midrule
    Resnet~\cite{hu2021two} & 0.746 & 0.710 & 0.736 & 0.784 \\
    ViT~\cite{koutini2021efficient} & 0.763 & - & - & - \\
    ViT~$\star$ & \textbf{0.763} & 0.739 & \textbf{0.770} & 0.779 \\
    PaSST~\cite{koutini2021efficient} & 0.756 & - & - & - \\
    PaSST~$\star$ & 0.756 & 0.740 & 0.737 & 0.791 \\
    FRITO-$r1$-$v8$~$\star$ & 0.761 & \textbf{0.745} & 0.735 & \textbf{0.804} \\
  \bottomrule
\end{tabular}
\end{center}
\end{table}

\subsection{Sparse Attention Efficiency}

Our Sparse FRITO-$r6$-$v1$ demonstrates a $20\%$ improvement in inference speed and a $21\%$ reduction in memory usage compared to its full-attention version, as outlined in Table~\ref{tab:efficiency}. We employ the same parameters as FRITO-$r6$-$v1$ on the Nsynth dataset for Sparse FRITO. However, it should be noted that the full-attention FRITO-$r6$-$v1$ experiences a $6\%$ degradation in inference speed compared to the vanilla ViT, which can be attributed to the overhead of computing the attention mask $M$.

\begin{table}[h]
\begin{center}
\caption{Comparison of inference speed and memory} \label{tab:efficiency}
\begin{tabular}{lcc}
  \toprule
  Method & Speed & Mem \\
  \midrule
    ViT~\cite{koutini2021efficient} & 0\% & 0\% \\ 
    FRITO-$r6$-$v1$ & -6\% & +0\% \\ 
    Sparse FRITO-$r6$-$v1$ & +14\% & -21\% \\ 
  \bottomrule
\end{tabular}
\end{center}
\end{table}

\section{Conclusion}

Our research is centered on enhancing the domain generalization capabilities of Transformer models in the sound classification task. By constraining the receptive field of the Transformer's self-attention, our model attains superior generalization performance on two publicly available datasets. Furthermore, we investigate sparse attention operations to optimize the model's inference speed and memory consumption.

Despite the empirical success, a theoretical foundation for the principle of limiting the receptive field remains to be established. Future work could endeavor to derive the relationship between the receptive field and generalization performance, providing a more rigorous understanding of the underlying mechanisms at play.



\bibliographystyle{IEEEtran}
\bibliography{citations}

\end{document}